\documentclass[11pt,preprint]{aastex}

\newcommand{\etal}{{\it et~al.}}

\begin{document}

\title{NEOWISE Reactivation Mission Year Three: Asteroid Diameters and Albedos}

\author{Joseph R. Masiero\altaffilmark{1}, C. Nugent\altaffilmark{2}, A.K. Mainzer\altaffilmark{1}, E.L. Wright\altaffilmark{3}, J.M. Bauer\altaffilmark{4}, R.M. Cutri\altaffilmark{2}, T. Grav\altaffilmark{5}, E. Kramer\altaffilmark{1}, S. Sonnett\altaffilmark{5}}

\altaffiltext{1}{Jet Propulsion Laboratory/California Institute of Technology, 4800 Oak Grove Dr., MS 183-301, Pasadena, CA 91109, USA, {\it Joseph.Masiero@jpl.nasa.gov}}
\altaffiltext{2}{California Institute of Technology, Infrared Processing and Analysis Center, 1200 California Blvd, Pasadena, CA 91125 USA}
\altaffiltext{3}{University of California, Los Angeles, CA, 90095, USA} 
\altaffiltext{4}{University of Maryland, College Park, MD, 20742, USA} 
\altaffiltext{5}{Planetary Science Institute, Tucson, AZ 85719 USA}

\begin{abstract}

The Near-Earth Object Wide-field Infrared Survey Explorer (NEOWISE)
reactivation mission has completed its third year of surveying the sky
in the thermal infrared for near-Earth asteroids and comets.  NEOWISE
collects simultaneous observations at $3.4~\mu$m and $4.6~\mu$m of
solar system objects passing through its field of regard.  These data
allow for the determination of total thermal emission from bodies in
the inner solar system, and thus the sizes of these objects.  In this
paper we present thermal model fits of asteroid diameters for $170$
NEOs and $6110$ MBAs detected during the third year of the survey, as
well as the associated optical geometric albedos.  We compare our
results with previous thermal model results from NEOWISE for
overlapping sample sets, as well as diameters determined through other
independent methods, and find that our diameter measurements for NEOs
agree to within $26\%$ (1-$\sigma$) of previously measured
values. Diameters for the MBAs are within $17\%$ (1-$\sigma$).  This
brings the total number of unique near-Earth objects characterized by
the NEOWISE survey to $541$, surpassing the number observed during the
fully cryogenic mission in 2010.

\end{abstract}

\section{Introduction}

The Near-Earth Object Wide-field Infrared Survey Explorer (NEOWISE)
reactivation mission is a NASA Planetary Science-funded survey using
an Earth-orbiting infrared telescope to detect and characterize
asteroids and comets in our solar system.  NEOWISE makes use of the
Wide-field Infrared Survey Explorer (WISE) spacecraft, which conducted
a 4-band thermal infrared survey of the entire sky from January 2010
until the exhaustion of cryogens in September 2010 \citep{wright10}.
Over the course of the primary mission, WISE detected over 150,000
asteroids and comets, characterizing their thermal infrared properties
and discovering over 30,000 new objects
\citep{mainzer11,mainzer11neo,masiero11,grav11,grav12,bauer13}.  The
mission was continued through February 2011 under the direction of
NASA's Planetary Science Division \citep{mainzer12,masiero12} at which
point the telescope was put into hibernation.  The telescope was
reactivated at the request of NASA's Planetary Science Division in
2013 to continue searching for new near-Earth objects (NEOs) and to
provide thermal infrared characterization of the observed NEOs
\citep{mainzer14}.  Oversight of NEOWISE was assumed by NASA's
Planetary Defense Coordination Office in 2016.

NEOWISE calibrated images and source detection catalog data are
released to the public annually through the NASA Infrared Processing
and Analysis Center (IPAC) Infrared Science Archive
(IRSA\footnote{\it{http://irsa.ipac.caltech.edu}}).  Concurrently, the
NEOWISE science team has published tables of derived physical
properties for the NEOs and Main Belt asteroids (MBAs) observed during
that year \citep[see][for Year 1 and Year 2 data,
  respectively]{nugent15,nugent16}.  Previously published physical
property data (up to NEOWISE Year 1) have been archived in the NASA
Planetary Data System \citep{mainzer16}; this archive will be updated
again at the end of the NEOWISE mission.  In this publication, we
present our derived physical properties from the NEOWISE Year 3
dataset.

\section{Observations and Followup}

NEOWISE is in a low-Earth, polar orbit, near the terminator.  Since
the exhaustion of cryogenic coolant in September 2010 the NEOWISE
detectors and telescope have been passively cooled via radiation to
deep space.  To facilitate this, NEOWISE is limited to pointing at
solar elongations larger than $\sim90^\circ$, and surveying near
zenith to minimize the heat-load from the Earth.  During the primary
mission zenith pointing coincided with $\sim90^\circ$ solar
elongation; however, as the orbit has precessed in the years following
launch the survey strategy has been modified to accommodate these
shifts.  On the side of the orbit precessing toward the Sun, the
telescope is actively pointed to scan at $90^\circ$ elongation at the
expense of additional heat load from the Earth, while on the other
side of the orbit the survey continues at zenith, drifting a few
degrees from the nominal elongation of $90^\circ$.  In addition to
these long-term changes, NEOWISE conducts toggles of a few degrees
during each quarter moon to avoid scanning over the moon and to
minimize the impact of lunar scattered light.  For more information,
refer to the NEOWISE Explanatory Supplement that is updated with each
annual data release \citep{cutri15}.

The NEOWISE telescope uses beamsplitters and co-aligned detectors to
simultaneously image the same $47^\prime\times47^\prime$ area of sky
onto two focal plane detectors with sensitivities centered at
$3.4~\mu$m and $4.6~\mu$m.  Each detector records the incident flux
for $7.7~$second exposures, followed by a $\sim2~$second slew of the
scan mirror that keeps the image stationary on the detectors during
exposures.  Exposures are separated by $11~$seconds, and have a
$\sim10\%$ overlap in the scan direction.  The nominal survey pattern
results in most detections of moving objects being spaced
$\sim3~$hours apart over a $\sim30~$hour period.

NEOWISE conducts a regular search of the survey data stream for known
and new moving objects through use of the WISE Moving Object
Processing System \citep[WMOPS,][]{mainzer11}.  Objects detected in a
minimum of 5 exposures are submitted to the IAU's Minor Planet Center
(MPC)\footnote{\it{http://www.minorplanetcenter.net}} for archiving.
The majority of objects detected by NEOWISE are recovery observations
of known objects linked by the MPC.  New objects with potential
NEO-like orbits are posted to the MPC's Near-Earth Object Confirmation
Page (NEOCP) for community followup, while new objects only having MBA
orbital solutions are archived and await future incidental followup.
The latter frequently have short orbital arcs (arc$<0.01$years) and
uncertain orbit solutions, and as such are excluded from physical
property analysis.

\begin{figure}[ht]
\begin{center}
\includegraphics[scale=0.6]{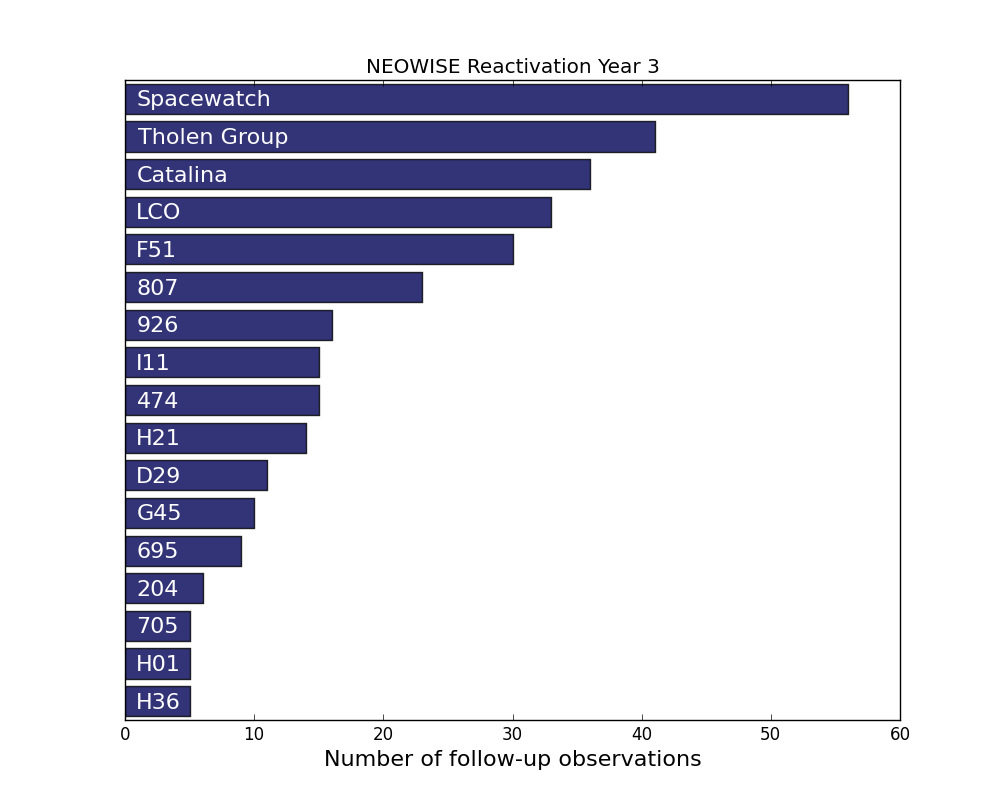}
\protect\caption{Followup observations of NEOWISE-discovered NEOs by
  ground-based facilities during the third year of survey.  Programs
  using multiple telescopes (the Spacewatch followup program, the
  followup program led by D. Tholen, the Catalina Sky Survey, and the
  Las Cumbres Observatory followup program) have been combined into
  single bins.  All other bins are identified by their Minor Planet
  Center observatory identification code.}
\label{fig.follow}
\end{center}
\end{figure}

For objects posted to the NEOCP, rapid followup observations from
telescopes around the world are critical to confirming the orbit and
determining the visible magnitude.  In Figure~\ref{fig.follow} we
present a record of the ground-based followup observations of NEOs
discovered by the NEOWISE survey in the third year of reactivation.
We note that the top seven followup programs are all funded by NASA's
Near-Earth Object Observations program as NEO discovery or followup
surveys, while the eighth (code I11) is a Gemini Large and Long
Program led by the first author to ensure recovery of faint NEO
candidates at southern declinations not observable by followup
telescopes in the northern hemisphere.

\section{Thermal Modeling Technique}

To determine asteroid physical properties from the measured infrared
flux values, we perform thermal modeling of the observed bodies using
their derived orbital parameters.  With our model of the thermal
behavior of the surface, we can constrain the diameter of the body by
comparing the predicted and measured thermal emission.  When optical
measurements are available in the literature or from ground-based
followup observations we also can constrain the albedo of the body.
During the NEOWISE mission only the $3.4\mu$m and
$4.6\mu$m channels (hereafter W1 and W2) are operational; measurements
at these wavelengths can only constrain the Wien side of the
blackbody emission, resulting in larger diameter uncertainties than
seen for data from the WISE/NEOWISE primary mission when longer
wavelength measurements ($12~\mu$m and $22~\mu$m) were available
\citep[c.f.][and discussion
  below]{mainzer12,masiero12,nugent15,nugent16}.  These previous
publications have detailed our thermal modeling technique, so here we
only present a synopsis of our methods, noting procedures that have
been updated.

\clearpage

\subsection{Data}

We extracted all detections from NEOWISE (observatory code C51)
recorded in the MPC's Observations
Catalog\footnote{\it{http://minorplanetcenter.net/iau/ECS/MPCAT-OBS/MPCAT-OBS.html}}
with observation dates between 13 December 2015 00:00 UT and 12
December 2016 23:59:59 UT.  This is done to provide a detection table
that has been vetted twice: once by the NEOWISE WMOPS system and a
second when linked by the MPC for orbit fitting.  WMOPS actively
removes stars and galaxies from the input detection list based on
real-time stationary detections in the data and the catalogs produced
from stacking of the cryogenic mission data.  This minimizes the
number of incorrect associations or bad detections included in our
thermal fitting.  These sets of Right Ascension-Declination-Time data
are used as input for a search of the NEOWISE Single-exposure
detection database hosted by the NASA/IPAC Infrared Science Archive
(IRSA).

Although this provides the best input dataset in terms of reliability,
it may lack in completeness, particularly for objects near the NEOWISE
detection threshold.  Objects near this limit that vary in flux due to
rotation will only be detected at the brightest points in their light
curve.  This will result in a potential over-estimation of the
diameters for faint objects.  For individual objects, it is possible
to search the single-exposure database at the predicted position for
the object in each frame and recover detections at SNRs below the
single-exposure limit of $4.5$ used by WMOPS for automated searches.
This more complete data set is critical for thermophysical modeling of
objects seen at multiple apparitions (e.g. 2015 QM$_3$; Wright et
al. 2017, in prep), but searching for additional low SNR data for all
the objects presented here is beyond the scope of this work.  However,
it is important to recognize the potential bias to larger sizes in our
fits of objects near the WMOPS detection limit.

In addition to extracting measured W1 and W2 magnitudes and errors,
the IRSA query of the NEOWISE Single-exposure detection database also
returns associated sources in the AllWISE 4-band Source Catalog within
$3''$.  This provides an extra level of filtering of static sources
beyond what is used by WMOPS.  We use these data to remove asteroid
detections that may be confused with background astrophysical objects,
such as faint stars and galaxies.  We remove any detection from our
input list that is coincident (within the search radius) to a
background object with SNR$~>7$ in either the W1 or W2 bands in the
AllWISE Catalog.  As the AllWISE detections are made on coadded images
of at least 8 Single-exposures this SNR cut corresponds to a flux
limit well below the single-exposure SNR of $4.5$ that WMOPS uses in
the search for moving objects.  However, we set a restriction that the
background object can be no more than $3~$magnitudes fainter (in the
AllWISE detection) than the detection associated with the asteroid;
this prevents faint background sources from triggering the loss of
valid detections of much brighter solar system objects.  Static
sources that show extreme brightening between the AllWISE Catalog and
the Year 3 observation will not be eliminated by the cut, but these
cases are expected to be rare and will not significantly affect the
results of our thermal modeling.

In order to mitigate the effects of cosmic rays striking very close to
the measured or predicted positions of asteroids in our images, we
also restricted the {\it w2rchi2} parameter of the detections to
$w2rchi2<5$ before carrying out our thermal fitting.  The $rchi2$
parameter measures the goodness-of-fit of the model PSF for each band
to the source extracted by the pipeline.  The mean value of {\it
  w2rchi2} for all detections is $0.95$, and $99.4\%$ of our detections
have $w2rchi2<5$ (before filtering).  W2 is the dominant thermal band
in all of our fits, and thus cosmic ray strikes contaminating this
band will lead to large errors in diameter.  As an example, Main Belt
asteroid (9190) was observed twice in Year 3, roughly five months
apart.  One detection in the second epoch was contaminated by a bright
cosmic ray strike, resulting in a spurious thermal model fit an order
of magnitude larger than both the other epoch and the diameter reported in
\citet{mainzer16}.  This single detection had a $w2rchi2=233.5$, while
the remaining detections were all near unity.  Because the cosmic ray
happened to strike the detector at nearly exactly the location
predicted for (9190), it was discarded by neither WMOPS nor the MPC,
but is eliminated by our filtering on the {\it w2rchi2} parameter.

To determine the heliocentric and geocentric distances at the time of
observation, we use the orbital elements for each object published by
the Minor Planet Center.  We restrict our input set to objects with
orbital arcs longer than $0.01$ years.  This is to ensure that the
orbit of the object is sufficiently constrained to produce accurate
distances.  This cut removes $3$ NEOs and $11$ MBAs from our fitting
list.  These objects received no optical followup, meaning the $\sim1$
day arc of observations from NEOWISE was the only positional data
available.

The positions of the spacecraft with respect to Earth for each
detection are drawn from the submitted data in the MPC Observations
Catalog.  Additionally, an estimate of the visible brightness at the
time of NEOWISE observations is required to determine a
visible-wavelength albedo.  We use the absolute $H$ magnitude and $G$
phase slope parameter to calculate this expected brightness.  When
available, we use the $H$ and $G$ values derived from the Pan-STARRS
photometric dataset by \citet{veres15} so long as they spanned
$>1^\circ$ of phase (a total of $5767$ objects).  Otherwise, we use
the $H$ and $G$ values published in the MPC orbital element file.
\citet{pravec12} highlighted specific systematic problems with the $H$
and $G$ determinations for some objects due to improper photometric
calibration of some surveys in the past. However recent photometry
from calibrated surveys (e.g. Pan-STARRS) has mitigated the majority
of these issues.

\begin{figure}[ht]
\begin{center}
\includegraphics[scale=0.6]{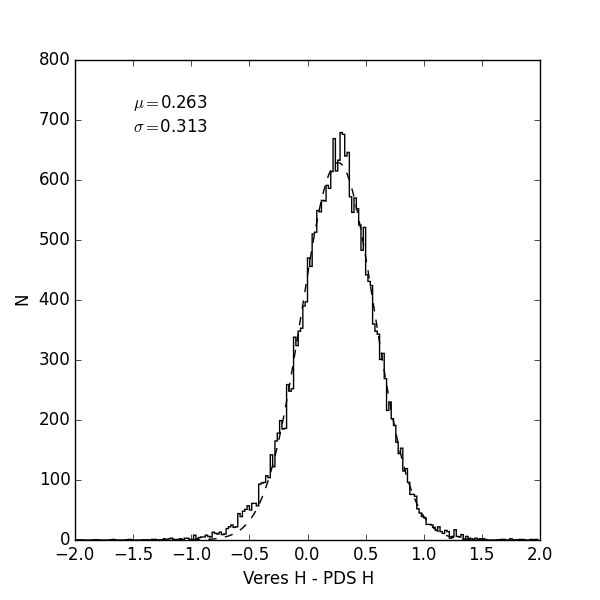}
\protect\caption{Histogram showing the difference in $H$ absolute
  magnitude for Main Belt asteroids from the MPC as used in
  \citet{mainzer16} compared with the updated $H$ values from
  \citet{veres15}.  The best fit Gaussian is shown as a dashed line,
  and the mean and standard deviation are given.}
\label{fig.Hdist}
\end{center}
\end{figure}

We note that there is a significant, systematic offset in the $H$
absolute magnitude of the asteroids in our sample as reported by
\citet{veres15} compared with the $H$ values drawn from the MPC used
for thermal fitting by previous work \citep[e.g.][]{masiero11}.  We
find an offset of $<\Delta H>=0.26~$mag for objects previously fit by
NEOWISE, as shown in Figure~\ref{fig.Hdist}, comparable to the
$<\Delta H>=0.22~$mag reported for the total population by
\citet{veres15}.  This offset will result in the predicted visual
brightness at the time of NEOWISE observation being $\sim27\%$ lower
on average when using the \citet{veres15} values, and thus the
calculated visible albedo being comparably lower.  We discuss further the
effects of this systematic change in $H$ below.

We assume an uncertainty of $0.05~$mag for $H$ for the Main Belt, and
$0.2~$mag for NEOs.  The latter is consistent with the observed
uncertainty in the MPC orbital catalog as well as typical asteroid
rotation amplitudes, however for some well-characterized bodies this
over-estimates the uncertainty.  Thus we use a smaller assumed $H$ error
for the Main Belt asteroids as this population is dominated by
low-numbered objects.  Our fitting routine simultaneously constrains
the NEOWISE-measured infrared magnitudes and the visible $H$ magnitude
by varying the diameter and albedo of the modeled object.  As the
optical flux has effectively one measurement ($H$) and one unknown
($p_V$ assuming diameter is constrained by the infrared), the majority
of fits reproduce the input $H$ magnitude exactly.  However, for cases
where the NEOWISE measurement has a very high SNR in both W1 and W2,
the least-squares minimizer can fall into a local minimum where the
reflected flux in W1 drives the optical albedo fit to unphysical
values.  For these cases, which tend to be bright asteroids with
well-constrained optical parameters, we assume an error on $H$ of
$0.05~$mag for the NEOs and $0.02~$mag for the MBAs.  This typically
succeeds in driving the fit out of the local minimum to the true
global best-fit.

\subsection{NEATM}

We apply the Near Earth Asteroid Thermal Model
\citep[NEATM][]{harris98} to a faceted sphere with the same orbital
state as the asteroid at the time of observation.  In order to account
for uncertainties in the thermophysical parameters of the asteroid
surface such as emissivity and conductivity, NEATM employs a variable
`beaming parameter' ($\eta$) to adjust the thermal behavior of the
surface from the idealized state.  This parameter acts as a sink
for uncertainties due to physical variation among objects as well as
random and systematic model errors.  NEATM is by necessity a simple
approximation of the real thermal behavior of asteroid surfaces; more
detailed thermophysical modeling
\citep[e.g.][]{spencer90,lagerros97,wright07,alilagoa14,koren15,hanus16,nugent17} can provide
improved constraints on the physical and thermal properties of an
asteroid if multiple viewing geometries are available, though at the
expense of greatly increased computation time.  We provide
NEATM-derived physical properties here as a springboard for more
detailed modeling of targets of interest.

Our general thermal fitting procedure is as follows: our routine reads
in the times, magnitudes, and spacecraft positions for each detection
returned from our search of the NEOWISE Year 3 data.  Detections in
the saturated non-linear regime \citep[c.f.][]{cutri15} are rejected,
as are any detections coincident with stationary sources in the
AllWISE catalog (see above).  The most recent orbit for the object is
read in from the Minor Planet Center's MPCORB file, and the position
of the object with respect to the sun and spacecraft is calculated for
each time of observation.  For each object, detections are clustered
into 10-day-long sets.  For distant objects (e.g. MBAs) this results
in multiple epochs of observations being fitted separately.  For NEOs
that follow the survey pattern and thus remain in the field of view
for long periods of time, this breaks up the single set of detections
into multiple subsets to accommodate any potential change in viewing
geometry over the course of the observations.  Our routine then
attempts to fit the diameter and albedo of each object with at least
three observations in one band, using assumed beaming parameters and
infrared-to-visible albedo ratios.

Our routine only uses detections with magnitude errors less than
$0.25$ (effectively restricting to brighter than SNR$\sim4$), will
only fit objects with at least 3 detections, and will only use the W1
band for fitting if the number of valid W1 detections is greater than
$40\%$ of the detections in W2.  During the fit, the visible and
infrared albedos are required to remain between $0.001$ and $1$ and
the beaming is required to remain between $0.3$ and $\pi$ (the
theoretical maximum for NEATM).  Based on this initial fit, our
routine checks the reflected and emitted light components in each
band.  If more than $75\%$ of the light in W1 is from reflected light,
the code reruns the fitting routine allowing the infrared albedo to be
a fitted parameter.  If instead both bands have at least three
detections and more than $25\%$ of the modeled flux based on the
initial fit is from thermal emission, the code reruns the fitting
routine allowing beaming parameter to be a fitted parameter.

These initial fits were then checked to verify that the modeled
visible flux matched the predicted flux based on the $H$ and $G$ phase
function parameters, that the visible albedo was larger than $0.01$
and smaller than $0.9$, and that at least one band had $\ge 90\%$ of its
modeled flux from thermal emission.  Fits that did not pass these
tests were rerun or dropped, as described below.

The $\eta$ values for inner solar system objects typically span the
range between $0.5-2.5$ \citep{mainzer11neo,masiero11}.  Lower values
result in a higher modeled subsolar temperature, leading to a larger
modeled flux per facet and a smaller modeled diameter.  Objects with
larger $\eta$ values have lower modeled subsolar temperatures and
larger modeled diameters.  During the cryogenic phase of the
WISE/NEOWISE survey, detections in multiple thermally-dominated bands
could be used to constrain the shape of an asteroid's spectral energy
distribution (SED), and thus the beaming parameter.  Using only the
$3.4$ and $4.6~\mu$m channels, only a handful of detected objects were
thermally dominated in both bands W1 and W2. Thus, an assumed value of
the beaming parameter had to be used for a majority of the detected
objects.  For objects that did have thermal emission in W1, W1 and W2
combined provide only a modest constraint on the shape of the SED as
both sample points that are close together on the blackbody curve. For
$7$ objects where the fitted beaming parameter resulted in unphysical
results we force the model to use the assumed beaming parameter (due
to an uncertain reflected light component in the W1 flux).  When
assuming a beaming value, we use the average value found for the NEOs
and MBAs from the cryogenic WISE/NEOWISE survey for each population:
$\eta_{NEO}=1.4\pm0.5$, $\eta_{MBA}=0.95\pm0.2$
\citep[][respectively]{mainzer11neo,masiero11}.  For a subset of
objects the initial NEATM fit resulted in an unnaturally low albedo
($p_V<0.01$), indicating that the beaming assumption had resulted in
an inaccurate diameter.  For these objects we use different assumed
beaming values of $\eta=0.8$ to obtain our best fit, and indicate them
in Tables~\ref{tab.NEOfits}-\ref{tab.MBAfits}.

In order to estimate the contribution of reflected light to the W1 and
W2 bands, we use an assumed infrared-to-optical geometric albedo ratio
of $p_{IR}/p_{V}=1.6\pm1.0$ for the NEOs and $p_{IR}/p_{V}=1.5\pm0.5$
for the MBAs.  These ratios are drawn from the average values for
those populations with $p_{IR}$ fitted using, at a minimum, the $3.4$ and
$12~\mu$m channels from the fully cryogenic WISE/NEOWISE data
\citep{mainzer11neo,masiero11}.  We assume that the albedos in W1 and
W2 are the same.  

To determine the statistical error component on our fitted parameters,
we perform 25 Monte Carlo trials of the fit, varying the three
measurements (H, W1, W2) by their respective error bars ($\sigma_H$,
$\sigma_{W1}$, $\sigma_{W2}$), while the assumed parameters (p$_{IR}$,
$\eta$) were varied by their estimated uncertainties.  In each trial,
the measured/assumed value is taken as the mean of a Gaussian, while
the uncertainty on each is used as the standard deviation; a new value
for each parameter is drawn randomly and independently from these
Gaussian distributions for each Monte Carlo trial.  Our thermal model
is then applied to these new input parameters, and the output best-fit
values are recorded.  Our quoted error on each parameter is the
standard deviation of the fitted values for that parameter in all of
the Monte Carlo runs.

In the trials, we vary the input infrared magnitude based on the
measurement error published in the single-exposure database.  Absolute
$H$ magnitude used error bar of $0.2~$mag for NEOs and $0.05~$mag for
MBAs, though as discussed above in some cases smaller errors were used
to reject solutions at a local minimum.  Assumed parameters were
assigned error values that were also varied according to their
uncertainties in the Monte Carlo trials.  Beaming parameter errors
were based on the spread of best fit beaming parameters found during
the cryogenic WISE/NEOWISE survey \citep{mainzer11neo,masiero11}.
Similarly, errors on the ratio between the infrared and visible albedo
were based on the range of fitted values in the cryogenic survey.
Listed errors for diameter and albedo thus represent the range of best
fit values for the Monte Carlo trials; these uncertainty values will
not include systematic errors due to the difference between the NEATM
model and the true thermal behavior of the asteroid.  Thus quoted
errors are a floor on the measurement error.

We note that due to a coding error in the tabulation of the
uncertainty determinations for the NEOWISE Year 1 and 2 data
\citep{nugent15,nugent16}, a subset of fits were reported to have
beaming uncertainties of $\sigma_\eta\sim0$, despite having assumed
beaming parameters and thus assumed uncertainty values.  In these
cases, the assumed $\sigma_\eta$ for each population (as above) was
indeed used for the Monte Carlo uncertainty determination but this
assumed uncertainty was not the value reported in the table.  We have
corrected this error in the current implementation of our thermal
fitting software; it does not alter the outcome of the Monte Carlo
determinations, and the only change to the previously published tables
is to replace the published uncertainties for the assumed beaming
parameters with the values describe in the text of those papers.  This
has been corrected for the results presented here.

\section{Results}

In Table~\ref{tab.NEOfits} we present $202$ NEATM-derived diameters
and albedos for $170$ unique NEOs.  Table~\ref{tab.MBAfits} gives
$6877$ NEATM-derived diameters and albedos for $6110$ unique MBAs.
These tables list the $H$ and $G$ values used as measurement inputs to
the fitting routine, best-fit diameter and albedo with associated
errors, the fitted or assumed beaming parameter (as indicated by the
final column, where $1$ indicates a fitted beaming and $0$ an assumed
value), the number of detections in each band used for the fit, and
the average phase angle of the observations used.  Objects listed
multiple times were detected at multiple apparitions, and were fit
independently.  This was done because non-spherical objects can have
different projected areas at different viewing geometries, even when
averaging over rotational phase.  Fits of different viewing geometries
that result in different diameter values may be statistical noise, or
may trace the true triaxial shape of the asteroid.  Objects with
different fitted diameters at multiple apparitions are candidates for
more sophisticated thermophysical modeling.

In Tables~\ref{tab.NEOfits} \& \ref{tab.MBAfits}, the format used to
present the best-fit albedos differs from previous years.  Here, we
provide the best-fit albedo as a base 10 logarithm, with errors on
that value.  We make this change because the errors associated with
our albedo determinations are inherently asymmetric and more
accurately captured in log-space.  Previously published albedo errors
were determined by the range of solutions from the Monte Carlo trials,
which naturally varied more in the positive direction than the
negative direction.  As such, some reported albedos appeared to have
errors that encompassed negative values.  These were accurate errors
for the positive direction, but overestimated the negative error.
This is corrected by use of log albedo, which shows that the errors
are symmetric around the mean as percentages of the best-fit.

\begin{figure}[ht]
\begin{center}
\includegraphics[scale=0.7]{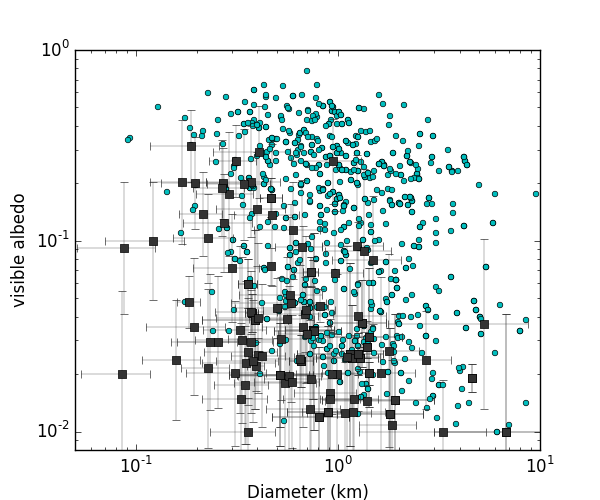}
\protect\caption{Comparison of fitted diameters and albedos for all
  near-Earth objects observed (cyan circles) and discovered (black
  squares) by NEOWISE during the first three years of the reactivation
  survey.  NEOWISE discoveries tend to be low albedo ($p_V<0.1$) and
  relatively large ($D>300~$m). Error bars on previously known objects
  are omitted for clarity, but are of comparable size to the errors on
  the NEOWISE discoveries.  }
\label{fig.diamalb}
\end{center}
\end{figure}

In Figure~\ref{fig.diamalb} we compare the diameters and albedos for
NEOs that were discovered by the NEOWISE during the first three years
of the reactivation survey to those of previously known NEOs that were
detected by NEOWISE during this same time-period.  NEOWISE
preferentially discovers NEOs with low albedos ($p_V<0.1$), filling in
the larger ($D>300~$m), dark objects that are missed by ground based
visible-light surveys due to their selection biases against low albedo
objects.  This includes $19$ NEOs with diameters measured to be larger
than 1 km.

During the process of verifying the thermal fits, we found that a
subset of objects showed unstable or unphysical solutions, and almost
all had in common a non-zero reflected light component to the
calculated W2 flux.  For objects that have a very high albedo and/or
are at large heliocentric distances at the time of detection, W2 may
include significant flux from both thermal and reflected light, and
the balance between these two components will depend very strongly on
the assumed beaming parameter; small changes to $\eta$ can result in
large changes in best-fit diameter.  To remove these potentially
erroneous fits, we discard any solution where the reflected light
component of the W2 flux was greater than $10\%$ of the total flux.
This resulted in $32$ NEOs and $4968$ MBAs that were observed by
NEOWISE during Year 3 not having fitted physical properties, and thus
are not included in Tables~\ref{tab.NEOfits} \&
\ref{tab.MBAfits}. This cut is more strict than was applied to the
NEOWISE reactivation Years 1 and 2 data but results in more robust
diameter determinations.  It is important to stress that this
restriction has a significant impact on the albedo distribution of the
MBAs reported here, causing a significant selection bias
\textbf{against} high albedo MBAs (as these will have a larger
contribution from reflected light at all wavelengths).  Thus low
albedo asteroids are over-represented in Table~\ref{tab.MBAfits}
compared to the results from the cryogenic survey ($90\%$ vs $62\%$
respectively).  This effect is less significant for NEOs as they tend
to be closer to the Sun at the time of observation, thus warmer and
with more emitted flux in W2.  Any attempt to characterize the whole
MBA population based on the fits from the Year 3 data will need to
account for this bias.  We note that this bias does not impact the
data from the cryogenic portion of the original WISE survey
\citep[see][]{mainzer11neo}, as the W3 and W4 fluxes were always entirely
from thermal emission.

\begin{table}[ht]
\begin{center}
\caption{Thermal model fits for NEOs in the third year of the
  NEOWISE survey.  Table 1 is published in its entirety
  in the electronic edition; a portion is shown here for guidance
  regarding its form and content.}
\vspace{1ex}
\noindent
\begin{tabular}{cccccccccc}
\tableline
  Name  &  H  &   G   &   Diameter  &  $\log{p_V}$  &   beaming  &  n$_{W1}$  &n$_{W2}$ & phase & Fitted \\
  & (mag) & & (km) &&&&& (deg) & Beaming? \\
\tableline
  01863 & 15.54 &  0.15 &    3.05 $\pm$   0.79 & -1.010 $\pm$ 0.297 & 2.12 $\pm$ 0.46 &  12 &  13 & 52.96 & 1\\
  01864 & 14.85 &  0.15 &    2.53 $\pm$   0.45 & -0.504 $\pm$ 0.142 & 1.27 $\pm$ 0.24 &   5 &   6 & 63.36 & 1\\
  01865 & 16.84 &  0.15 &    0.71 $\pm$   0.26 & -0.319 $\pm$ 0.267 & 1.40 $\pm$ 0.53 &   5 &   5 & 69.40 & 1\\
  02100 & 16.05 &  0.12 &    3.05 $\pm$   0.37 & -1.242 $\pm$ 0.317 & 2.56 $\pm$ 0.26 &  32 &  33 & 60.30 & 1\\
  02102 & 16.00 &  0.15 &    1.76 $\pm$   0.60 & -0.654 $\pm$ 0.256 & 1.58 $\pm$ 0.48 &  22 &  27 & 53.60 & 1\\
  03360 & 15.90 &  0.15 &    2.56 $\pm$   1.14 & -0.930 $\pm$ 0.341 & 1.40 $\pm$ 0.50 &   0 &   8 & 40.98 & 0\\
  03360 & 15.90 &  0.15 &    2.34 $\pm$   0.35 & -0.848 $\pm$ 0.121 & 1.25 $\pm$ 0.17 &   6 &   6 & 58.47 & 1\\
  04179 & 15.30 &  0.10 &    1.79 $\pm$   0.38 & -0.392 $\pm$ 0.166 & 1.01 $\pm$ 0.22 &   6 &   6 & 47.91 & 1\\
  04341 & 16.11 &  0.10 &    3.18 $\pm$   0.65 & -1.264 $\pm$ 0.427 & 2.65 $\pm$ 0.47 &   6 &   6 & 49.48 & 1\\
  04769 & 16.90 &  0.15 &    1.40 $\pm$   0.03 & -1.036 $\pm$ 0.062 & 2.72 $\pm$ 0.05 &   5 &   5 & 61.91 & 1\\

\hline
\end{tabular}
\label{tab.NEOfits}
\end{center}
\end{table}

\begin{table}[ht]
\begin{center}
\caption{Thermal model fits for MBAs in the third year of the
  NEOWISE survey.  Table 2 is published in its entirety
  in the electronic edition; a portion is shown here for guidance
  regarding its form and content.}
\vspace{1ex}
\noindent
\begin{tabular}{cccccccccc}
\tableline
  Name  &  H  &   G   &   Diameter  &  $\log{p_V}$  &   beaming  &  n$_{W1}$  &n$_{W2}$ & phase & Fitted \\
  & (mag) & & (km) &&&&& (deg) & Beaming? \\
\tableline
  00013 &  6.74 &  0.15 &  216.06 $\pm$  54.78 & -1.313 $\pm$ 0.198 & 0.95 $\pm$ 0.20 &  12 &  12 & 24.55 & 0\\
  00019 &  7.13 &  0.10 &  205.42 $\pm$  68.11 & -1.339 $\pm$ 0.249 & 0.95 $\pm$ 0.20 &   6 &   8 & 24.27 & 0\\
  00034 &  8.51 &  0.15 &  113.26 $\pm$  30.34 & -1.412 $\pm$ 0.295 & 0.95 $\pm$ 0.20 &  13 &  14 & 21.84 & 0\\
  00035 &  8.60 &  0.15 &  126.62 $\pm$  41.41 & -1.482 $\pm$ 0.246 & 0.95 $\pm$ 0.20 &   8 &   9 & 24.42 & 0\\
  00035 &  8.60 &  0.15 &  140.67 $\pm$  47.03 & -1.537 $\pm$ 0.251 & 0.95 $\pm$ 0.20 &  10 &  11 & 24.91 & 0\\
  00036 &  8.46 &  0.15 &  106.13 $\pm$  32.38 & -1.360 $\pm$ 0.285 & 0.95 $\pm$ 0.20 &  13 &  13 & 22.52 & 0\\
  00041 &  7.12 &  0.10 &  184.78 $\pm$  60.20 & -1.286 $\pm$ 0.320 & 0.95 $\pm$ 0.20 &   9 &   9 & 23.54 & 0\\
  00049 &  7.65 &  0.19 &  151.30 $\pm$  39.27 & -1.329 $\pm$ 0.233 & 0.95 $\pm$ 0.20 &  11 &  11 & 22.11 & 0\\
  00050 &  9.24 &  0.15 &   87.92 $\pm$  27.53 & -1.428 $\pm$ 0.368 & 0.95 $\pm$ 0.20 &   7 &   7 & 22.83 & 0\\
  00056 &  8.31 &  0.15 &  109.09 $\pm$  34.20 & -1.273 $\pm$ 0.237 & 0.95 $\pm$ 0.20 &   6 &   6 & 26.66 & 0\\
\hline
\end{tabular}
\label{tab.MBAfits}
\end{center}
\end{table}

\clearpage

\section{Comparison of Derived Properties to Literature Values}

To quantify how accurately our modeled diameters reflect the actual
effective spherical diameter of the asteroids observed, we compare the
fits presented in this paper with diameter measurements of the same
asteroids in three independent data sets.  The first dataset is
composed of a collection of diameters derived from radar reflection
measurements
\citep{hudson94,magri99,shevchenko06,magri07,shepard10,naidu15,benner17},
occultation timing chords \citep{durech11,dunhamOcc}, and {\it in
  situ} spacecraft measurements, as discussed in \citet{mainzer11cal}.
These techniques do not rely on thermal modeling to derive diameters,
and thus allow us to probe any systematic issues with our thermal
modeling technique.  However, these calibrators are also the smallest
in number of our three comparison sets, and mainly consist of only the
largest asteroids.

Our second comparison set consists of objects with diameters fitted
using data from the fully cryogenic portion of the original
WISE/NEOWISE survey where the beaming parameter was a fitted value.
As the $12~\mu$m and $22~\mu$m bandpasses were the most sensitive to
thermal emission from asteroids, they typically provided high SNR
measurements and allowed for the beaming parameter $\eta$ to be
fitted.  These diameter measurements use the same modeling techniques,
allowing us to determine the effect of the assumed beaming parameter
on the overall fit quality.  Comparison diameters were obtained from
the Planetary Data System (PDS) archive \citep{mainzer16}, which is a
compilation of data from \citet{mainzer11neo} and
\citet{masiero11,masiero14}.

Our final comparison set is the collection of objects that were observed
during Years 1 or 2 of the NEOWISE mission, as well as in
Year 3.  As these fits were conducted using nearly identical methods
and assumptions, this comparison allows us to determine the effect of
changing viewing geometry on the overall fit quality.  Comparison
diameters were obtained from the \citet{mainzer16} PDS archive which
included fits from \citet{nugent15} as well as the fits presented by
\citet{nugent16} of asteroids observed in the NEOWISE Year 2 data.

\begin{figure}[ht]
\begin{center}
\includegraphics[scale=0.5]{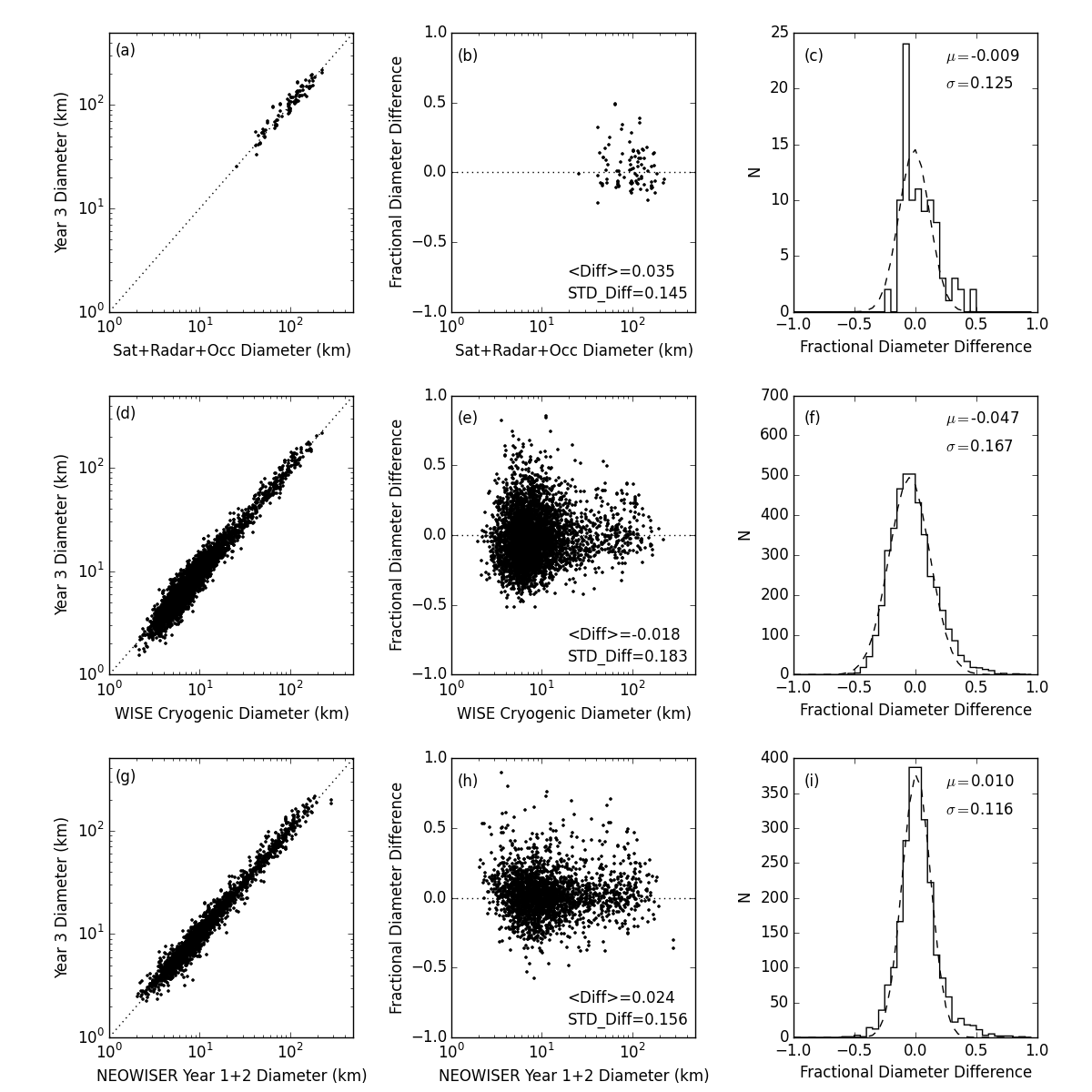} 

\protect\caption{Main Belt asteroid diameter fits from the NEOWISE
  Year 3 data compared to diameters derived from satellite, radar, and
  occultation measurements (panel (a)), NEOWISE fully cryogenic data
  (panel (d)), and NEOWISE-R Year 1 and 2 data (panel(g)).  Dotted
  lines show a 1:1 relationship.  We show the fractional difference in
  fits against the comparison diameter ((year 3 -
  comparison)/comparison; panels (b), (e), (h)) for each comparison
  set.  We also show the histogram of the fractional differences
  (panels (c), (f), (i)) along with the best-fit Gaussian to the
  fractional difference distribution and its mean ($\mu$) and standard
  deviation ($\sigma$).}
\label{fig.mbacomp}
\end{center}
\end{figure}

Figure \ref{fig.mbacomp} shows the comparison for the Main Belt
asteroids of the diameters in the three independent sets with the
diameters presented in this work.  We present two different
statistical assessments of the diameters we fit in this work.  In
panels (b), (e), and (h) we give the numerical mean ``$<$Diff$>$'' and
standard deviation ``STD\_DIFF'' of the fractional diameter
differences between the Year 3 diameter and the comparison data.  In
panels (c), (f), and (i) we fit a Gaussian distribution to the
histogram of the fractional diameter differences.  This fit is shown
as a dashed line, with the fitted $\mu$ and $\sigma$ of the Gaussian
shown in the each panel.  We prefer the fitted Gaussian for our
analysis as it is less affected by small numbers of outliers than the
simple mean calculation, however we list both for completeness.  From
the comparison to satellite, radar, and occultation diameters, the
largest objects have a diameter uncertainty $1-\sigma$ spread of
$12.5\%$ based on a Gaussian fit to the diameter differences.  The
overall diameter uncertainty is $\sigma<17\%$ for all three comparison
data sets.  The systematic offset in mean diameter difference is at
most a few percent for all cases.

\begin{figure}[ht]
\begin{center}
\includegraphics[scale=0.5]{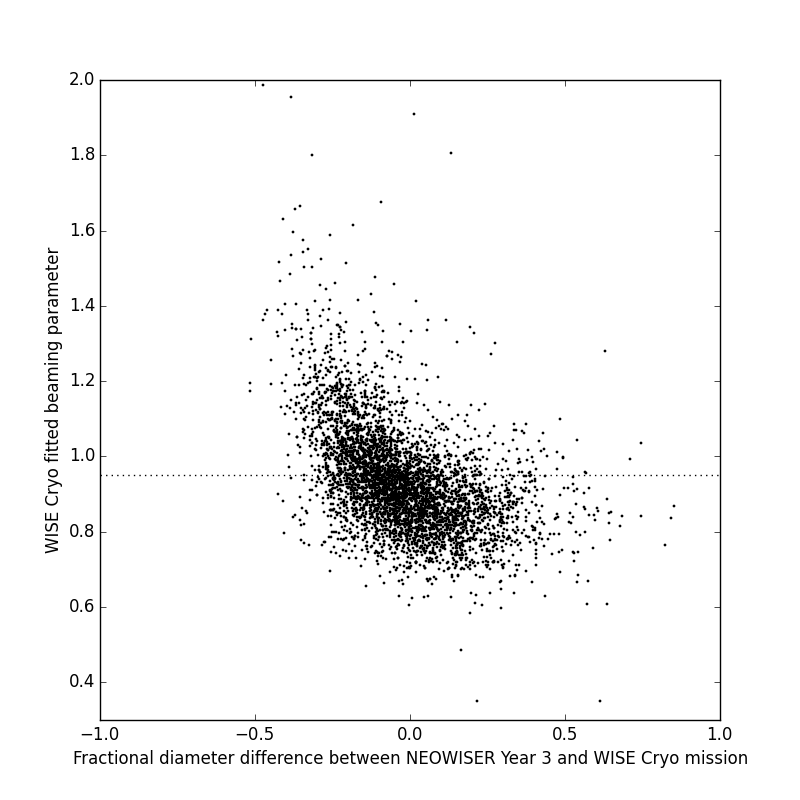}
\protect\caption{Comparison of the difference between the MBA fits
  presented in this paper and MBA fits from the fully cryogenic
  portion of the WISE mission showing the non-linear and asymmetric
  effect of beaming on this difference.  Although the distribution of
  beaming parameters has a long tail to values larger than our assumed
  $\eta=0.95$ (dotted line), the majority of objects have values lower
  than our assumed value, resulting in a general trend to positive
  fractional diameter differences.}
\label{fig.beamscat}
\end{center}
\end{figure}

All three comparisons show an asymmetry in the fractional diameter
difference, where the Year 3 diameters are skewed to larger sizes.
This is a result of our assumed value for the beaming parameter.  As
shown in Figure~\ref{fig.beamscat}, our assumed beaming value is near
the mean for the population, but the difference between this value and
the value measured during the cryogenic mission affects the diameter
error in a non-random way.  A true beaming value (that would be fit to
an ideal data set fully sampling the spectral energy distribution)
that is larger than our assumed one would result in a modeled peak
surface temperature higher than the actual
value ($T_{model}\propto\frac{1}{\eta^{0.25}}$) and thus a smaller
diameter than would be found in the ideal case.  Conversely, an actual
beaming smaller than our assumption would result in a systematically
larger fitted diameter value.  Although beaming varies over a larger
range above $\eta=0.95$, the majority of objects observed during the
Year 3 survey had previously measured beaming values below
$\eta=0.95$, resulting in the asymmetric distributions seen in
Fig~\ref{fig.mbacomp}.  We note, however, that the uncertainty on the
assumed beaming parameter is captured in our Monte Carlo simulations
used to derive the statistical diameter error, so much of this effect
is represented in the quoted errors.

\begin{figure}[ht]
\begin{center}
\includegraphics[scale=0.5]{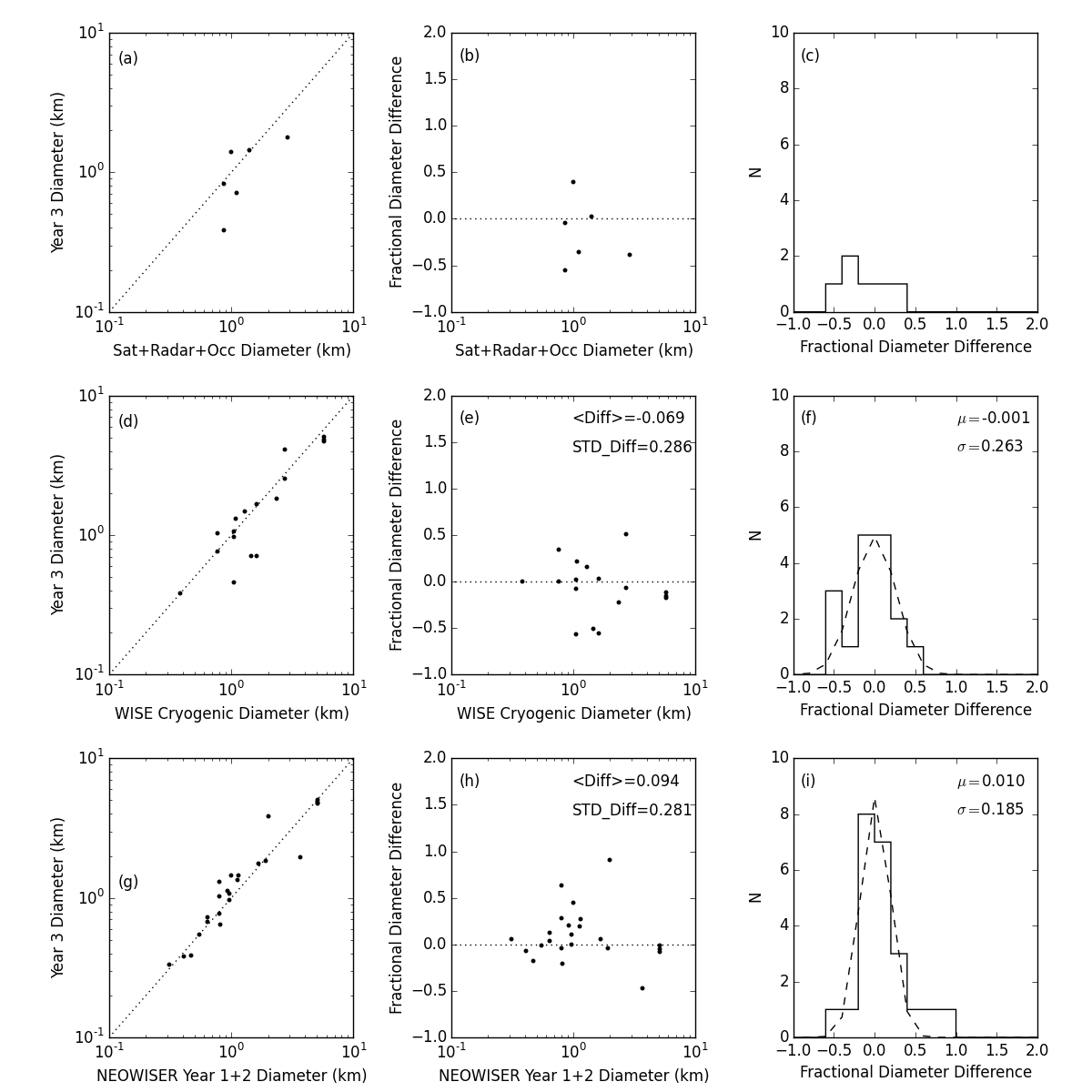}
\protect\caption{The same as Figure~\ref{fig.mbacomp}, but for
  near-Earth objects observed during Year 3 that also were present in
  one of the comparison datasets.}
\label{fig.neocomp}
\end{center}
\end{figure}

We show in Figure~\ref{fig.neocomp} the same comparison for NEOs.
There are only a small number of overlapping objects in the satellite,
radar, and occultation data set, making it impossible to draw robust
conclusions from this comparison and thus we do not present the
statistical metrics for this set.  Comparison with the cryogenic
NEOWISE fits shows an overall $1-\sigma$ diameter uncertainty of
$26\%$, with a smaller dispersion for the comparison to previous years
of NEOWISE reactivation data.  As with the MBAs, the mean offset in
for NEO diameters is less than a few percent. The diameter offset is
larger when looking at the numerical mean of the fractional
differences for both comparison sets, but the limited sample size
makes this metric particularly sensitive to a small number of
outliers.

\begin{figure}[ht]
\begin{center}
\includegraphics[scale=0.5]{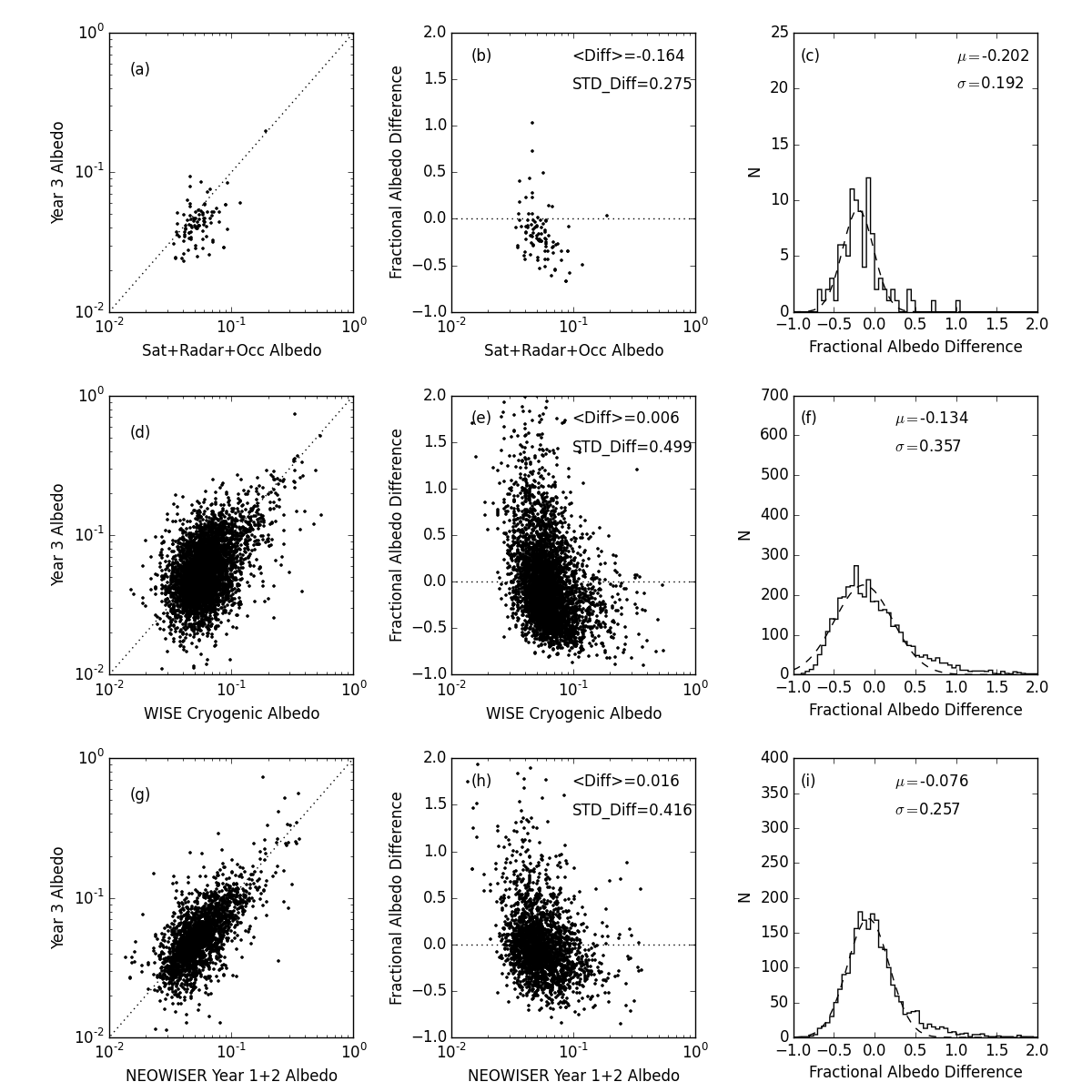} 

\protect\caption{Main Belt asteroid albedo fits from the NEOWISE Year
  3 data compared to albedos derived from satellite, radar, and
  occultation measurements (panel (a)), NEOWISE fully cryogenic data
  (panel (d)), and NEOWISE-R Year 1 and 2 data (panel(g)).  Dotted
  lines show a 1:1 relationship.  We show the fractional difference in
  fits against the comparison albedo ((year 3 -
  comparison)/comparison; panels (b), (e), (h)) for each comparison
  set.  We also show the histogram of the fractional differences
  (panels (c), (f), (i)) along with the best-fit Gaussian to the
  fractional difference distribution and its mean ($\mu$) and standard
  deviation ($\sigma$).  The large mean offsets for all three
  comparison sets are a result of the use of updated $H$ values in the
  fits presented here, which systematically increased the $H$ values
  for a large number of asteroids, and thus lowered the fitted
  albedo. }
\label{fig.mbacomp_alb}
\end{center}
\end{figure}

We also can compare the albedos we derive from our fits of the Year 3
data to the same calibration sets as for the diameters. We show this
comparison in Figure~\ref{fig.mbacomp_alb}.  There is a strong
systematic bias to lower albedos in our fits of the Year 3 data as
compared to previous fits of the same objects.  As discussed above,
this is a result of the systematic offset in the $H$ absolute
magnitudes published by \citet{veres15} as compared to the MPC-derived
$H$ magnitudes used for the thermal model fits by
\citet{masiero11,masiero14,nugent15}.  The MBA fits from
\citet{nugent16} for NEOWISE Year 2 used $H$ magnitudes from
\citet{williams12}, who revised the MPC photometric fits.  A
comparison of the $H$ magnitudes from \citet{veres15} and
\citet{williams12} to those from the MPC used in the NEOWISE PDS
release show a similar offset to fainter absolute magnitudes for both
sets.  This explains why the comparison of albedos from Year 3 to
those from Years 1 and 2 shows a smaller offset than the comparison to
the cryogenic data.  As the albedos given for the objects with
diameter measurements from satellite, radar, and occultations were
also based older $H$ magnitude fits, they will also show an offset
when compared to albedos derived from the updated $H$ values.  As
Figure~\ref{fig.mbacomp_alb} highlights, revisions to the literature
absolute magnitudes can result in significant changes to the albedos
determined from thermal modeling. However as shown in
Figure~\ref{fig.mbacomp} these changes in $H$ magnitude have no
significant impact on the accuracy of the diameter determination.

\section{Conclusions}

We have presented NEATM thermal model fits for NEOs and MBAs detected
by the NEOWISE mission during its third year of
surveying.  In total, there were $170$ NEOs and $6110$ MBAs with
sufficient data to constrain a diameter based on their thermal
emission, and an albedo based on literature reflected visible light.
We find that our Main Belt asteroid diameter fits have $1-\sigma$
uncertainties of $<20\%$ when compared to other published diameters.
The NEOs have a larger uncertainty of $<30\%$.  However this is based
on a small number of objects.  For the NEOWISE Year 3
data, we have tightened our selection requirements, rejecting any fits
where more than $10\%$ of the modeled flux in W2 was from reflected
light.  This most severely impacted the number of Main Belt asteroids
with published fits (rejecting nearly $5000$ objects).  However, we
feel this improves the reliability of the remaining fits.  This cut
does introduce a strong bias against high albedo objects in the Main
Belt, which are more likely to have a significant amount of reflected
light in W2, resulting in the majority of our reported Main Belt
asteroids being from the low-albedo complex.

NEOWISE is continuing its survey for asteroids and comets into its
fourth year.  Orbital precession will eventually make surveying
impossible at some point following the end of the fourth year of
survey, though this depends on the activity level of the Sun.  It
remains difficult to predict when conditions will inhibit high-quality
data collection, but until that time the data remain highly useful for
the characterization and discovery of near-Earth asteroids, Main Belt
asteroids, and comets.

\clearpage

\section*{Acknowledgments}

This publication makes use of data products from the Wide-field
Infrared Survey Explorer, which is a joint project of the University
of California, Los Angeles, and the Jet Propulsion
Laboratory/California Institute of Technology, funded by the National
Aeronautics and Space Administration.  This publication also makes use
of data products from NEOWISE, which is a project of the Jet
Propulsion Laboratory/California Institute of Technology, funded by
the Planetary Science Division of the National Aeronautics and Space
Administration.  This research has made use of data and services
provided by the International Astronomical Union's Minor Planet
Center.  This publication uses data obtained from the NASA Planetary
Data System (PDS).  This research has made use of the NASA/IPAC
Infrared Science Archive, which is operated by the Jet Propulsion
Laboratory, California Institute of Technology, under contract with
the National Aeronautics and Space Administration.  Based on
observations obtained at the Gemini Observatory through a Gemini Large
and Long Program, which is operated by the Association of Universities
for Research in Astronomy, Inc., under a cooperative agreement with
the NSF on behalf of the Gemini partnership: the National Science
Foundation (United States), the National Research Council (Canada),
CONICYT (Chile), Ministerio de Ciencia, Tecnolog\'{i}a e
Innovaci\'{o}n Productiva (Argentina), and Minist\'{e}rio da
Ci\^{e}ncia, Tecnologia e Inova\c{c}\~{a}o (Brazil).  Finally, the
authors acknowledge the efforts of NEO followup observers around the
world who provide time-critical astrometric measurements of newly
discovered NEOs, enabling object recovery and computation of orbital
elements.  Many of these efforts would not be possible without the
financial support of the NASA Near-Earth Object Observations Program.

\end{document}